\documentclass{PoS}
\usepackage{graphics}
\newcommand{\be}{\begin{eqnarray}}
\newcommand{\ee}{\end{eqnarray}}
\newcommand{\bfn}{\begin{figure}[htb]\begin{center}}
\newcommand{\efn}{\end{center}\end{figure}}
\newcommand{\btn}{\begin{table}[htb]\begin{center}}
\newcommand{\etn}{\end{center}\end{table}}
\newcommand{\av}[1]{\langle #1 \rangle}
\newcommand{\mapright}[1]{\smash{\mathop{\hbox to 1cm{\rightarrowfill}}\limits^{#1}}}
\newcommand{\lw}[1]{\smash{\lower 1.5ex\hbox{#1}}}

\title{Analyses of multiplicity distributions and Bose-Einstein correlations at the LHC by means of generalized Glauber-Lachs formula}

\ShortTitle{generalized Glauber-Lachs formula}

\author{\speaker{Takuya Mizoguchi}\\
        Toba National College of Maritime Technology, Toba 517-8501, Japan\\
        E-mail: \email{mizoguti@toba-cmt.ac.jp}}

\author{Minoru Biyajima\\
        Department of Physics, Shinshu University, Matsumoto 390-8621, Japan\\
        E-mail: \email{biyajima@azusa.shinshu-u.ac.jp}}

\abstract{Using the negative binomial distribution (NBD) and the generalized Glauber-Lachs (GGL) formula, we analyze the data on charged multiplicity distributions in the several pseudorapidity intervals $|\eta| < \eta_c$ at 0.2 - 7 TeV by UA5 and ALICE Collaborations. We confirm that the KNO scaling holds among the multiplicity distributions with $\eta_c =$ 0.5 at $\sqrt{s} =$ 0.2 - 2.36 TeV and estimate the energy dependence of a parameter $1/k$ in NBD and parameters $1/k$ and $\gamma$ (the ratio of the average value of the coherent hadrons to that of the chaotic hadrons) in the GGL formula. Using empirical formulae for the
parameters $1/k$ and $\gamma$ in the GGL formula, we predict the multiplicity distributions with $\eta_c =$ 0.5 at 7 and 14 TeV. Data on the second order Bose-Einstein correlations (BEC) at 0.9 and 2.36 TeV by ALICE and CMS Collaborations are also analyzed based on the GGL formula. Predictions for the third order BEC at 0.9 and 2.36 TeV are presented.}

\FullConference{The Seventh Workshop on Particle Correlations and Femtoscopy\\
		 September 20 - 24 2011\\
		 University of Tokyo, Japan}

\begin{document}

\vspace{-2mm}
\section{Introduction}
\label{se_01}
\vspace{-2mm}
Recently ALICE Collaboration~\cite{Aamodt:2010ft} has investigated the multiplicity distributions with pseudo-rapidity cutoffs and compared its data with the data by UA5 Collaboration \cite{Alner:1985zc,Ansorge:1988kn}, and concluded that the combined data with $\eta_c =$ 0.5 at 0.2, 0.9, and 2.36 TeV are fairly well described by the single NBD (negative binomial distribution)~\cite{GrosseOetringhaus:2009kz,Fuglesang:1990aa}. Moreover, ALICE Collaboration has reported that the KNO scaling~\cite{Koba:1972ng} holds among the combined data with $\eta_c =$ 0.5 at 0.2, 0.9, and 2.36 TeV. The first aim of this study is to confirm the statement above mentioned in~\cite{Aamodt:2010ft} and to analyze the same data by the GGL(generalized Glauber-Lachs) formula~\cite{Biyajima:1982un,Biyajima:1983qu}. Some predictions at 2.36 TeV~\cite{Aamodt:2010pp} are also included in this proceeding.

Moreover, ALICE and CMS Collaborations have reported the data on Bose-Einstein correlations (BEC)~\cite{Aamodt:2010jj,Khachatryan:2010un}. Thus we investigate them based on a conventional formula with the degree of coherence and the GGL formula. According to main results in Ref.~\cite{Mizoguchi:2010vc}, our talk is presented.

The NBD is introduced in the following:
\be
P_k(n) = \frac{\Gamma(n+k)}{\Gamma(n+1)\Gamma(k)} 
     \frac{(\av{n}/k)^n}{(1+\av{n}/k)^{n+k}},
\label{eq_01}
\ee
where $\av{n}$ and $k$ are the average multiplicity and the intrinsic parameter, respectively. In the KNO scaling limit ($n$ and $\av{n}$ are large, but the ratio $z = n/\av{n}$ is finite), for the quantity $\av{n}P(n,\:\av{n})$ the following gamma distribution is derived from Eq.~(\ref{eq_01}) as
\be 
\psi_k(z) = \frac{k^k}{\Gamma(k)}z^{k-1}e^{-kz}
\label{eq_02}
\ee
Second we turn to the GGL formula which is expressed as follows:
\be
P_k(n) = \frac{(p\av{n}/k)^n}{(1+p\av{n}/k)^{n+k}}
     \exp\left[-\frac{\gamma p\av{n}}{1+p\av{n}/k}\right]
      L_n^{(k-1)}\left(-\frac{\gamma k}{1+p\av{n}/k}\right),
\label{eq_03}
\ee
where $\gamma = |\zeta|^2/A$ (the ratio of the average value of the coherent hadrons to that of the chaotic hadrons), $p = 1/(1+\gamma )$, and $L_n^{(k-1)}$ stands for the Laguerre polynomials, respectively. 

The KNO scaling function of Eq.~(\ref{eq_03}) is given in the following
\be 
\psi_k(z,\:p) = \left(\frac{k}{p}\right)^k 
        \left[\frac{z}{\sqrt{z(k/p)^2(1-p)}}\right]^{k-1}
        \!\!\!\!\!\! \exp\left[-\frac{k}{p}(1-p+z)\right] 
        I_{k-1}\left(2\sqrt{z(k/p)^2(1-p)}\right)
\label{eq_04}
\ee
where $I_{k-1}$ is the modified Bessel function. Eq.~(\ref{eq_04}) becomes the gamma distribution, as $\gamma =$ 0.

In order to analyze of Bose-Einstein correlations (GGLP effect~\cite{Goldhaber:1960sf}, or hadronic HBT effect~\cite{Biyajima:1990ku,Weiner:1997kg}) at LHC, we are going to use the following formulae: The first one is well known as the conventional formula, 
\be
&&N^{(--)}/N^{BG} ({\rm conventional\ formula})
= c[1 + \lambda\: E_{2B}^2],
\label{eq_05}\\
&& N^{(--)}/N^{BG} ({\rm GGL})
= c[1 + 2p(1-p)E_{2B} + p^2E_{2B}^2]
\label{eq_06}
\ee
where $c$ is normalization factor, $\lambda$ is the degree of coherence, $p = 1/(1 + \gamma)$ and $E_{2B}$ is function of momentum transfer ($Q^2 = -(p_1-p_2)^2$) and the range of interaction $R$. $E_{2B} = \exp(-R^2Q^2)$ (Gaussian formula) and/or $E_{2B} = \exp(-R\sqrt{Q^2})$ (exponential formula) are used. 

\vspace{-2mm}
\section{Analyses of data on multiplicity distributions by the NBD and the GGL formula}
\label{se_02}
\vspace{-2mm}
Utilizing Eqs.~(\ref{eq_01}) and (\ref{eq_03}), we analyze the data with pseudo-rapidity cutoffs ($\eta_c =$ 0.5, 1.0, and 1.3) at 0.2, 0.54, 0.9 and 2.36 TeV. Results at 0.9 and 2.36 TeV are shown in Fig.~\ref{fi_01}.
\bfn
\vspace{-2mm}
\resizebox{0.42\textwidth}{!}{\includegraphics{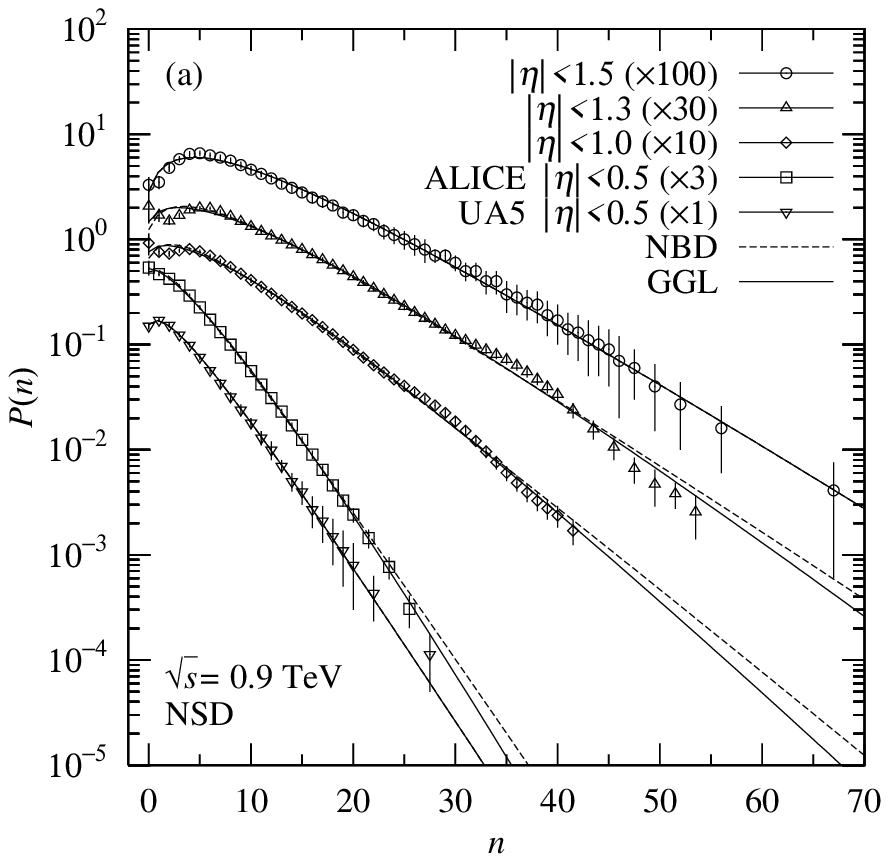}}
\resizebox{0.42\textwidth}{!}{\includegraphics{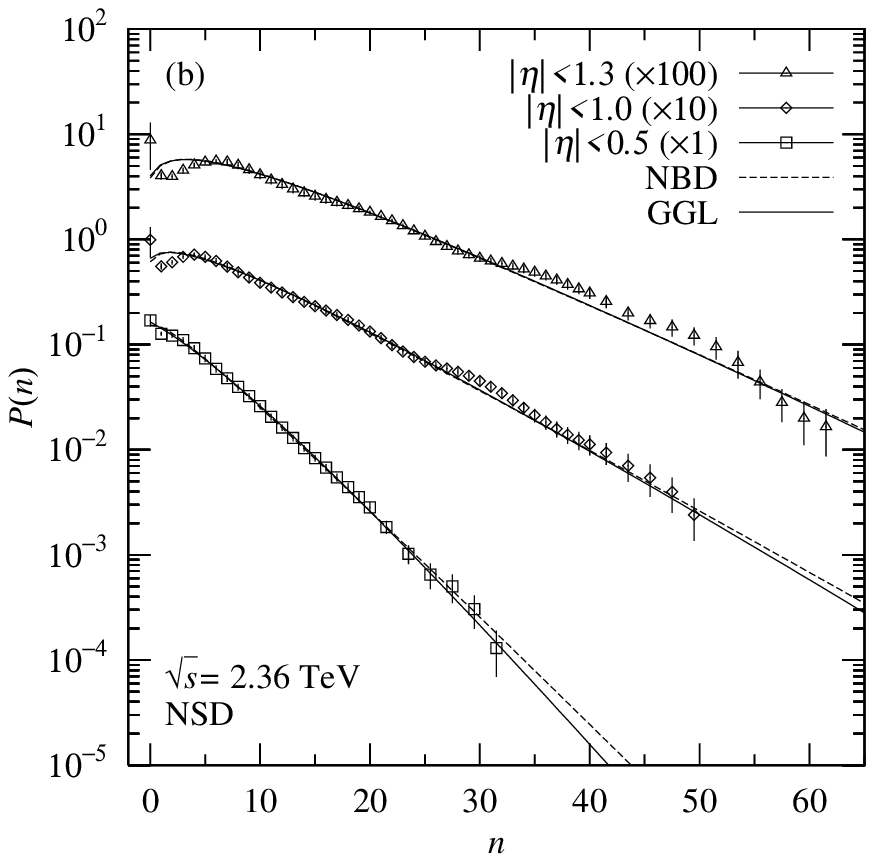}}
\vspace{-4mm}
\caption{Analyses of data with $|\eta|<\eta_c$ by means of Eqs.~(\protect \ref{eq_01}\protect) and (\protect \ref{eq_03}\protect). (See estimated values of parameters contained in Eqs.~(\protect \ref{eq_01}\protect) and (\protect \ref{eq_03}\protect) shown in Fig.~\protect \ref{fi_02} and \cite{Mizoguchi:2010vc}).}
\label{fi_01}
\vspace{-6mm}
\efn
Energy dependences of parameters $1/k^{\rm (NBD)}$, $1/k^{\rm (GGL)}$ and $\gamma^{\rm (MD)}$ (MD: multiplicity distribution) with $\eta_c =$ 0.5 are shown in Fig.~\ref{fi_02}. We observe that $1/k^{\rm (NBD)}$ increases gradually as $\sqrt{s}$ increases. On the other hand, the estimated sets of ($1/k^{\rm (GGL)}$ and $\gamma^{\rm (MD)}$) in the GGL formula show different behavior. 
\bfn
\vspace{-2mm}
\resizebox{0.50\textwidth}{!}{\includegraphics{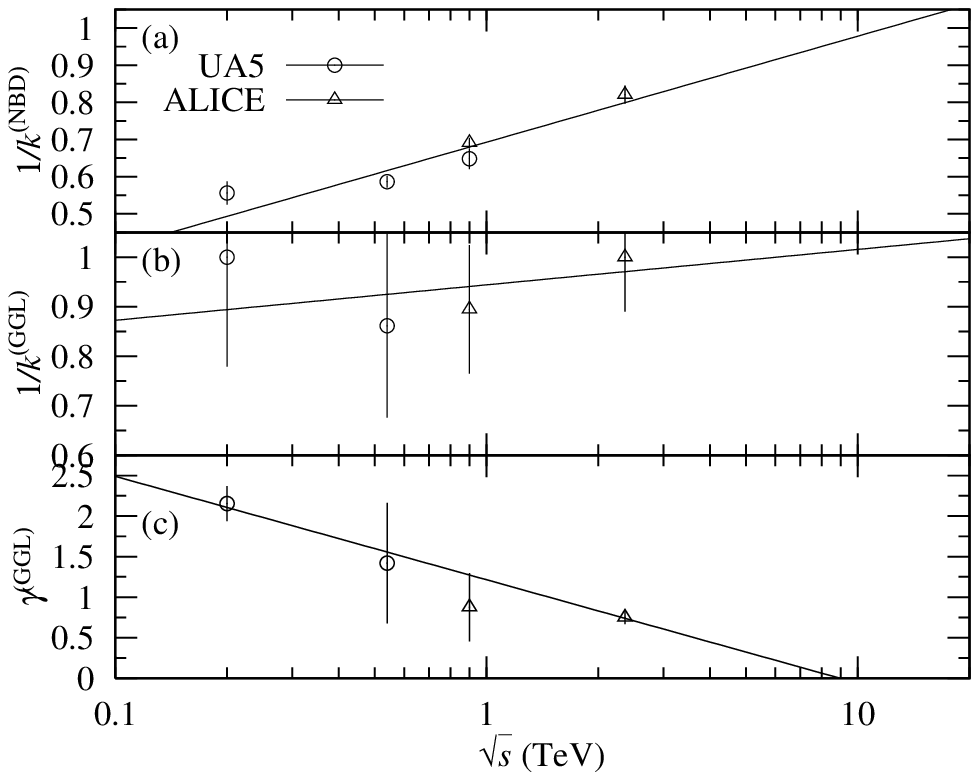}}
\label{fi_02}
\vspace{-4mm}
\caption{Energy dependences of parameters $1/k^{\rm (NBD)}$, $1/k^{\rm (GGL)}$, and $\gamma^{(MD)}$ for data with $\eta_c =$ 0.5. Values at 0.9 TeV by UA5 Collaboration are omitted in GGL formula, because of extreme error bars.}
\vspace{-6mm}
\efn

\vspace{-2mm}
\section{Analyses of data on KNO scaling distributions by Eqs.~(\protect \ref{eq_02}\protect) and (\protect \ref{eq_04}\protect)}
\label{se_03}
\vspace{-2mm}
Utilizing the KNO scaling variable $z (= n/\av{n})$, data on the KNO scaling distributions $\av{n} P(n,\: \av{n})$ are shown in Fig.~\ref{fi_03}. 
\bfn
\vspace{-2mm}
\resizebox{0.42\textwidth}{!}{\includegraphics{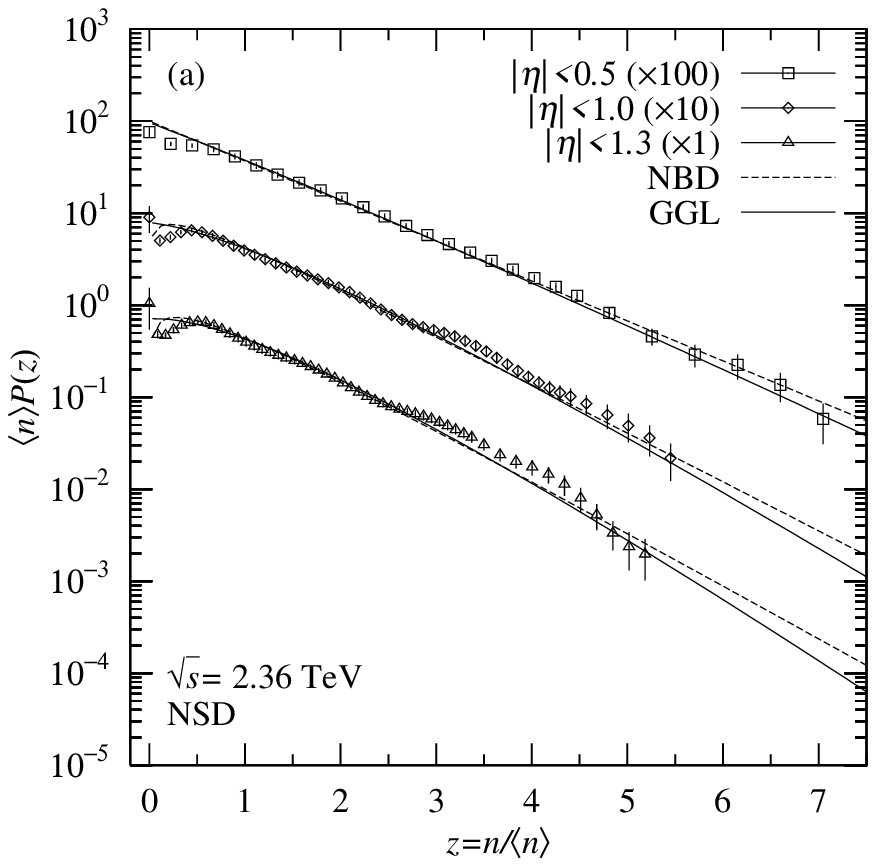}}
\resizebox{0.42\textwidth}{!}{\includegraphics{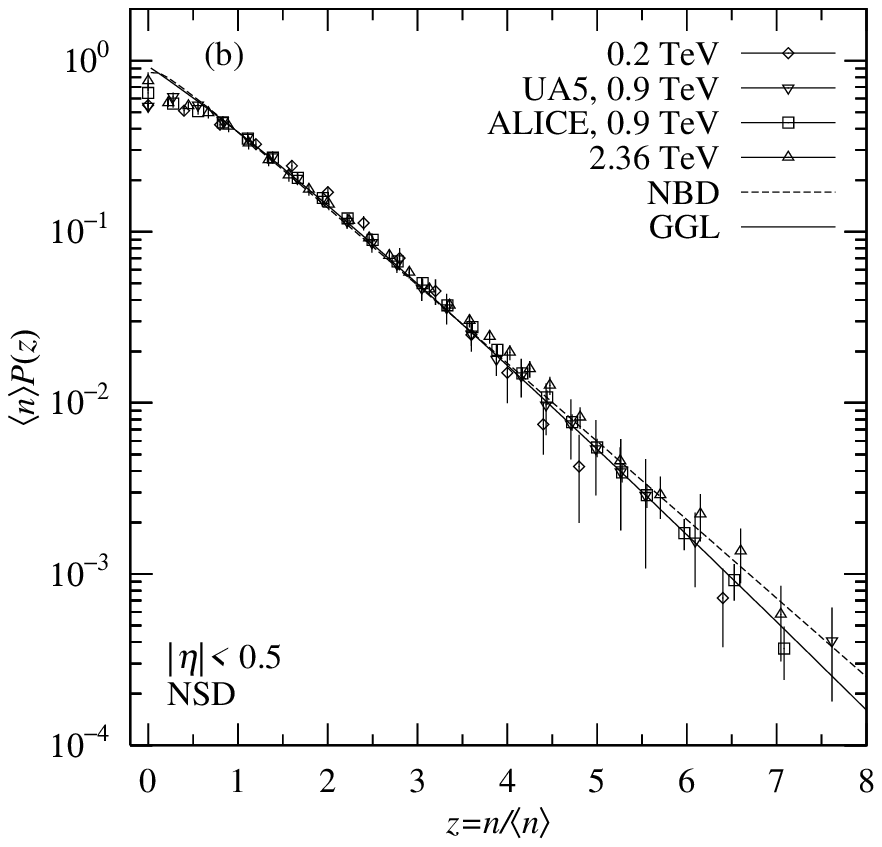}}
\vspace{-4mm}
\caption{Analyses of KNO scaling distributions $\av{n}P(n)$'s. The same data of Fig.~\protect \ref{fi_01}\protect  are described by KNO scaling variable $z=n/\av{n}$. Eqs.~(\protect \ref{eq_02}\protect) and (\protect \ref{eq_04}\protect) are used.}
\label{fi_03}
\vspace{-6mm}
\efn
We combine the data with $\eta_c =$ 0.5 at 0.2, 0.9 and 2.36 TeV and analyze them by Eq.~(\ref{eq_02}) (the gamma distribution) and Eq.~(\ref{eq_04}) (the modified Bessel function). 

\vspace{-2mm}
\section{Analyses of data on the 2nd order BEC by means of Eqs.~(\protect \ref{eq_05}\protect) and (\protect \ref{eq_06}\protect)}
\label{se_04}
\vspace{-2mm}
We analyze the data on BEC at LHC by the use of Eqs.~(\ref{eq_05}) and (\ref{eq_06}) with $E_{2B} = \exp(-R^2Q^2)$ and/or $E_{2B} = \exp(-R\sqrt{Q^2})$. Results are depicted in Table~\ref{ta_04} and Fig.~\ref{fi_04}. In Eq.~(\ref{eq_06}), the effective degree of coherence ``$\lambda$'' is ``$(1 + 2\gamma)/( 1 + \gamma)^2$''. In our concrete analyses, we obtained that $1/k^{\rm (BEC)} =$ 1. $\gamma^{\rm (BEC)}$ is similar to the value at 0.9 TeV by ALICE Collaboration in Fig.~\ref{fi_02}. 

Furthermore, by the use of Eqs.~(\ref{eq_05}) and (\ref{eq_06}) with $c$, we have analyzed the data on BEC at 0.9 and 2.36 TeV by CMS Collaboration~\cite{Khachatryan:2010un}. Results are shown in Fig.~\ref{fi_04} and Table~\ref{ta_04}. Notice that estimated values of $\lambda$, $\gamma$ and $R$ do not depend on the range of exclusive region ($0.4 < Q < 1.4$ GeV/$c$). It is emphasized that the ratio $\gamma^{\rm (BEC)}$ decreases, as the colliding energy increases. In other words, the effective degree of coherence ``$\lambda$'' and the range of interaction $R$ increases from 0.9 to 2.36 TeV. To draw more significant meaning about the parameter $\gamma$, we need BEC measurements with $\eta_c =$ 0.5.

\bfn
\vspace{-2mm}
\resizebox{0.41\textwidth}{!}{\includegraphics{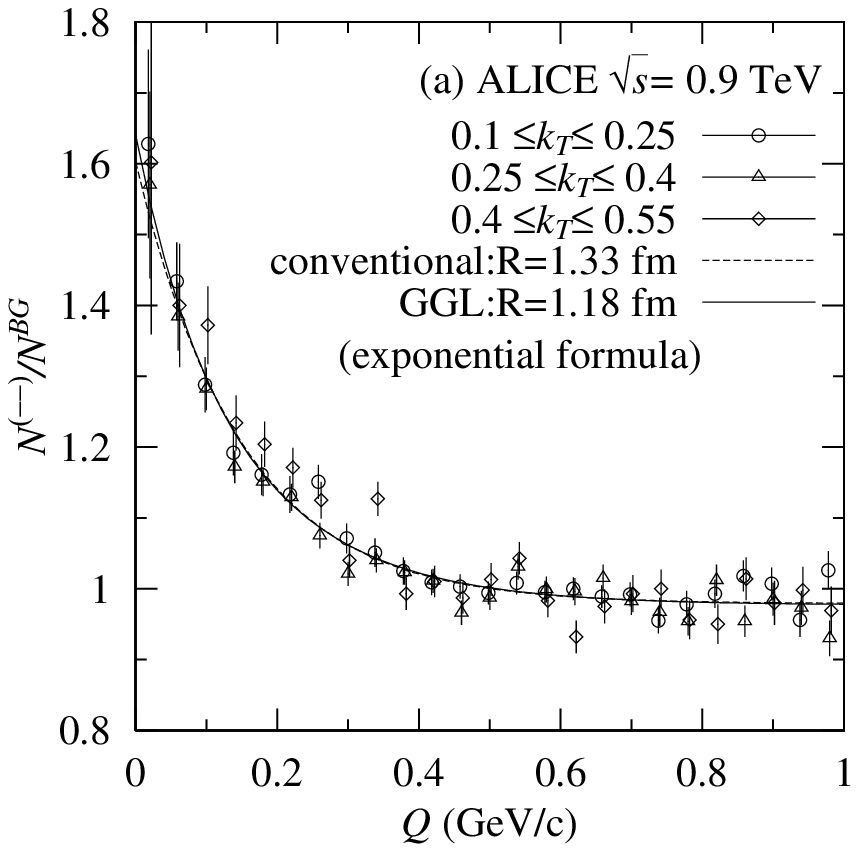}}
\resizebox{0.57\textwidth}{!}{\includegraphics{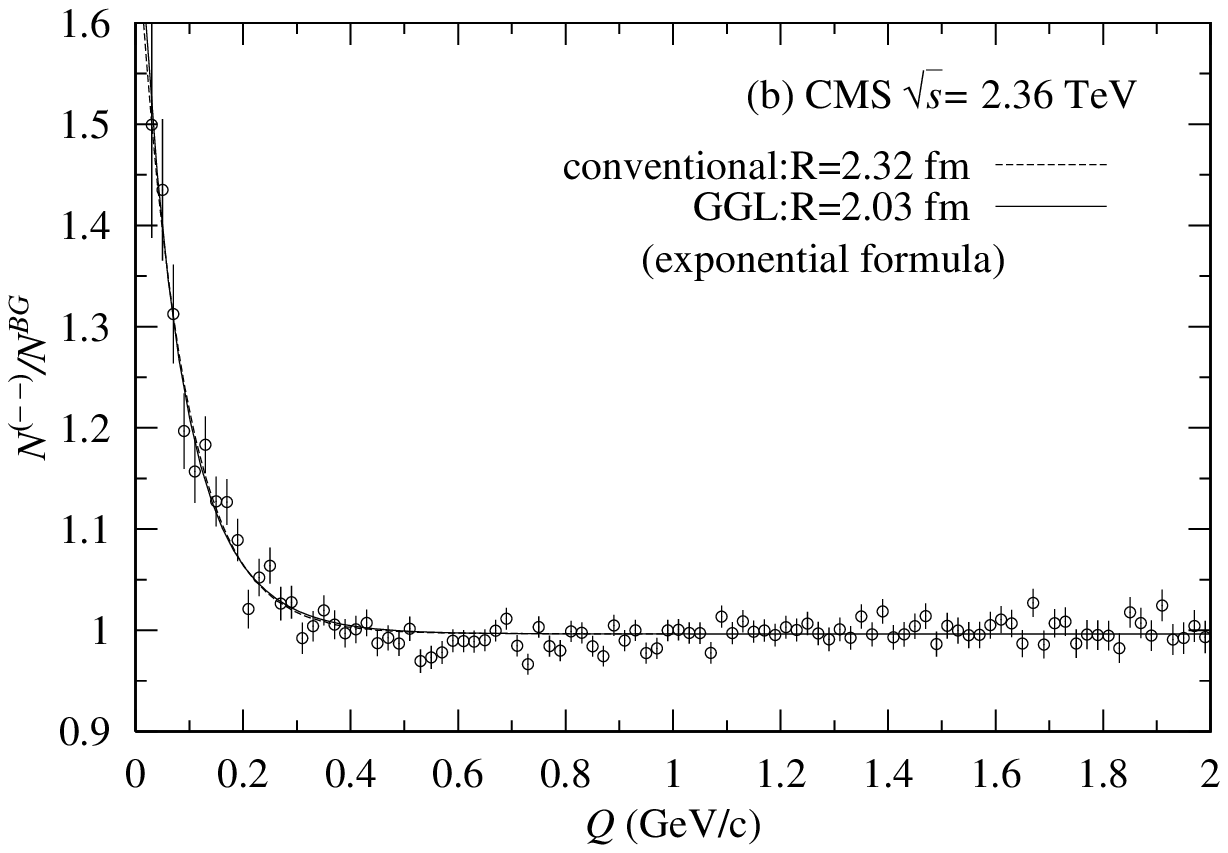}}\\
\vspace{-4mm}
\caption{Analyses of data on BEC at 0.9 TeV by ALICE Collaboration with conditions $M\le$ 6, and 0.1 $\le k_T \le$ 0.55 GeV and at 2.36 TeV by CMS Collaboration.}
\label{fi_04}
\vspace{-6mm}
\efn

\btn
\vspace{-3mm}
\caption{Analysis of data on BEC by ALICE Collaboration and CMS Collaboration. Because estimated value of $1/k^{(\rm BEC)}$ is a unit, it is not cited.}
\vspace{2mm}
\label{ta_04}
\begin{tabular}{cccc|cccc}
\hline
\multicolumn{4}{c|}{Eq.~(\ref{eq_05})} & \multicolumn{4}{c}{Eq.~(\ref{eq_06})}\\
\multicolumn{8}{c}{(upper: Gaussian formula, and lower: exponential formula)}\\
$\lambda$ & $c$ & $R$ (fm) & $\chi^2/$NDF & 
$\gamma$  & $c$ & $R$ (fm) & $\chi^2/$NDF\\
\hline
\multicolumn{8}{c}{$\sqrt{s} =$ 0.9 TeV, ALICE (multiplicity $M\le$ 6, 0.1 $\le k_T \le$ 0.55 GeV)}\\
\hline
0.35$\pm$0.02 & 0.988$\pm$0.003 & 0.83$\pm$0.04 & 121/72 & 
4.0$\pm$ 0.3  & 0.988$\pm$0.003 & 0.81$\pm$0.03 & 119/72\\
0.64$\pm$0.04 & 0.979$\pm$0.004 & 1.33$\pm$0.09 & 98/72 &
1.30$\pm$0.23 & 0.977$\pm$0.004 & 1.18$\pm$0.07 & 98/72\\
\hline
\multicolumn{8}{c}{$\sqrt{s} =$ 0.9 TeV, CMS (Excluding 0.6 $<Q<$ 0.9 GeV/$c$)}\\
\hline
0.32$\pm$0.01 & 0.995$\pm$0.001 & 0.96$\pm$0.02 & 407/165 & 
4.5$\pm$ 0.2  & 0.995$\pm$0.001 & 0.95$\pm$0.02 & 394/165\\
0.66$\pm$0.02 & 0.993$\pm$0.001 & 1.75$\pm$0.04 & 229/165 & 
1.08$\pm$0.13 & 0.993$\pm$0.001 & 1.59$\pm$0.03 & 225/165\\
\hline
\multicolumn{8}{c}{$\sqrt{s} =$ 2.36 TeV, CMS (Excluding 0.6 $<Q<$ 0.9 GeV/$c$)}\\
\hline
0.33$\pm$0.03 & 0.997$\pm$0.002 & 1.20$\pm$0.07 & 80/81 & 
4.3$\pm$ 0.6  & 0.997$\pm$0.002 & 1.18$\pm$0.07 & 80/81\\
0.72$\pm$0.08 & 0.997$\pm$0.002 & 2.32$\pm$0.17 & 75/81 & 
0.84$\pm$0.39 & 0.996$\pm$0.002 & 2.03$\pm$0.08 & 76/81\\
\hline
\end{tabular}
\vspace{-6mm}
\etn

\bfn
\resizebox{0.45\textwidth}{!}{\includegraphics{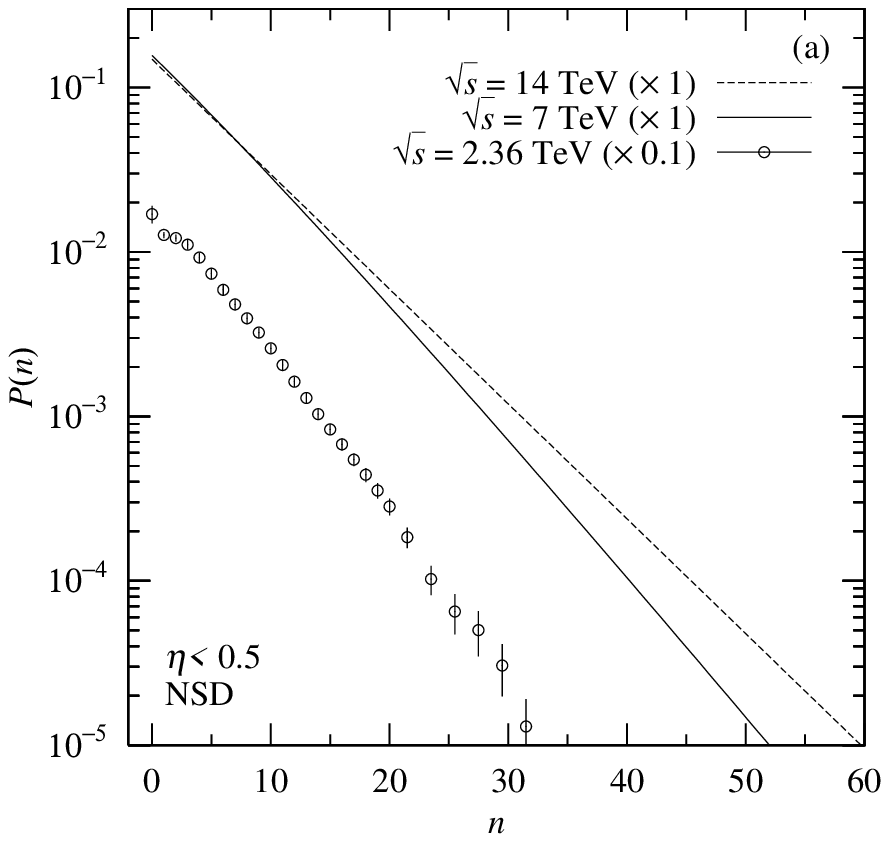}}
\resizebox{0.45\textwidth}{!}{\includegraphics{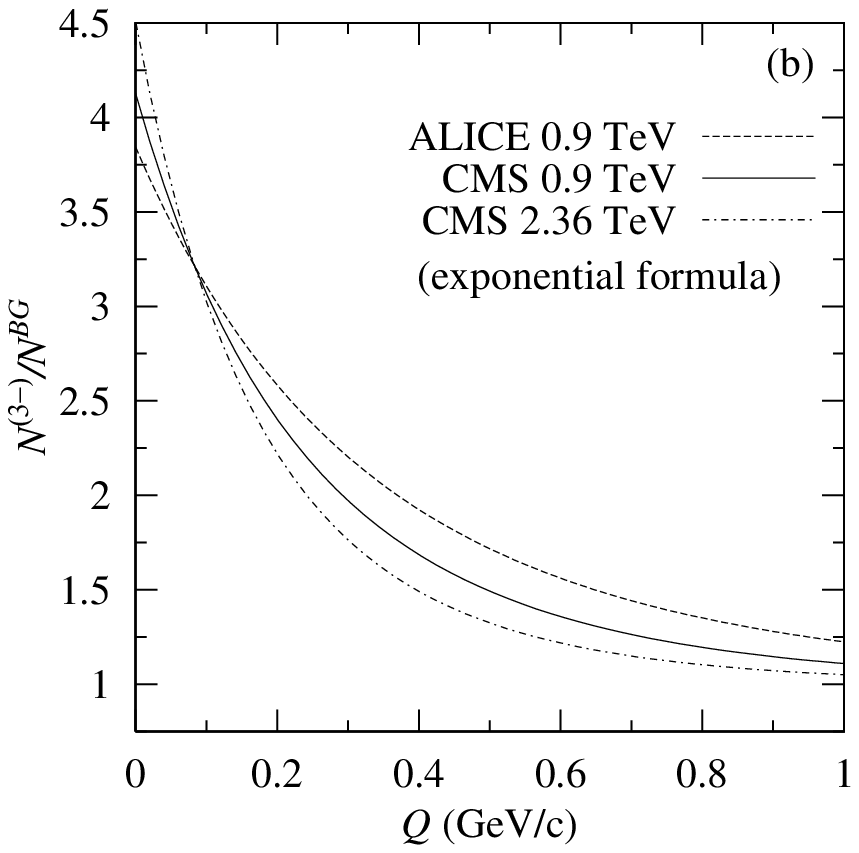}}
\vspace{-4mm}
\caption{(a) Expected multiplicity distributions with $\eta_c =$ 0.5 at $\sqrt{s} =$ 7 and 14 TeV. Computations are based on the GGL formula (Eq.~(\protect \ref{eq_03}\protect)) with values in Fig.~\protect \ref{fi_02} and $\av{n} = 2.5+0.76\ln(\sqrt{s}/0.2)$. (b) Our predictions of the 3rd order BEC at 0.9 and 2.36 TeV. Eq.~(\protect \ref{eq_15}\protect) with values in Table\protect \ref{ta_04} is used.}
\label{fi_05}
\vspace{-6mm}
\efn

\vspace{-2mm}
\section{Concluding remarks}
\label{se_05}
\vspace{-2mm}
We have confirmed that the multiplicity distributions with $\eta_c =$ 0.5 are described by the single NBD~\cite{Aamodt:2010ft}. Moreover, we also confirm that the GGL formula does work well for the explanation of the same data in present analyses.

We observed that distributions with $\eta_c =$ 0.5 at 7 TeV does not have the coherent component. In other words, the multiplicity distributions with $\eta_c =$ 0.5 at 7 TeV are described by the NBD with $k=$ 1.

Using values in Fig.~\ref{fi_02}, we can predict multiplicity distributions with $\eta_c =$ 0.5 at 7 and 14 TeV in Fig.~\ref{fi_05}a. Those are able to be examined in a near future. If there were discrepancies among data and predictions, we should consider the other effect, for example, due to the mini-jets~\cite{Giovannini:1998zb}.

Through present analyses of the BEC, results by the exponential formula seem to be better than those by the Gaussian formula in Table~\ref{ta_04}. See~\cite{Shimoda:1992gb} for the source functions. Moreover, values of $\gamma$'s obtained in Fig.~\ref{fi_02} and Table~\ref{ta_04} seem to be similar each other. To obtain more significant knowledge on the parameter $\gamma$, analyses of the multiplicity distributions and the BEC in the same hadronic ensembles are necessary\cite{Biyajima:1990ku,Weiner:1997kg}. 

It is worthwhile to predict the 3rd BEC at 0.9 TeV using the same condition with $M\le$ 6, 0.1 $\le k_T \le$ 0.55 GeV/$c$. Utilizing estimated values of $\gamma^{\rm (BEC)}$ and $R$ in the 2nd BEC by ALICE Collaboration, we can predict the 3rd order BEC; The following formula~\cite{Biyajima:1990ku} is used,
\be
N^{(3-)}/N^{BG}
= 1 + 6p(1-p)e^{-\frac{1}{3}R\sqrt{Q_3^2}}
+ 3p^2(3-2p)e^{-\frac{2}{3}R\sqrt{Q_3^2}}
+ 2p^3e^{-R\sqrt{Q_3^2}},
\label{eq_15}
\ee
where $p=1/(1+\gamma)$ and $Q_3^2 = Q_{12}^2 + Q_{23}^2 + Q_{31}^2$. Our predictions on the 3rd order BEC at 0.9 and 2.36 TeV are given in Fig.~\ref{fi_05}b. The results would be compared with measurements, as UA1 Minimum Bias Collaboration did~\cite{Neumeister:1991bq}. By these comparisons, we could obtain more useful information on the parameter $\gamma$ and the role of the GGL formula.\medskip

\noindent
{\bf Addendum:}\quad Recently CMS Collaboration has reported new analyses on BEC at 0.9 and 7 TeV in $pp$ collisions~\cite{Khachatryan:2011hi}. We have applied Eq.~(\ref{eq_06}) to data with the exponential form and the long range effect $(1+\alpha Q)$. $R=1.47$ fm (0.9 TeV) and $R=1.8$ fm (7 TeV) are obtained~\cite{Mizoguchi:2012aa}.

\end{document}